\documentclass[
aps,
pra,
twocolumn,
superscriptaddress,
showpacs,
english,
10pt,
floatfix
]
{revtex4-2}

\usepackage[utf8]{inputenc}
\usepackage{physics}
\usepackage[nohyperlinks,nolist]{acronym}
\usepackage{graphicx}
\usepackage[dvipsnames]{xcolor}
\usepackage{microtype}
\usepackage{bm}
\usepackage{url}
\usepackage{blindtext}
\usepackage{chemformula}
\usepackage[caption=false]{subfig}
\usepackage[version=4]{mhchem}
\usepackage{here}

\usepackage[linktocpage=true,
  colorlinks=true, 
  pdfborder={0 0 0},
  linkcolor=blue,
  citecolor=red,
  filecolor=yellow,
  urlcolor=blue,
  bookmarks,
  pdfauthor={},
]{hyperref}
\usepackage[capitalise]{cleveref}

\newcommand{\UniBasel}{Department of Physics, University of Basel, Klingelbergstrasse 82, CH-4056 Basel, Switzerland}
\newcommand{\rub}{Lehrstuhl f\"ur Theoretische Chemie II, Ruhr-Universit\"at Bochum, 44780 Bochum, Germany}
\newcommand{\rccs}{Research Center Chemical Sciences and Sustainability, Research Alliance Ruhr, 44780 Bochum, Germany}

\newcommand{\partition}{\omega_i(\vb r)}
\newcommand{\abinit}{{\textit{ab initio}}}
\newcommand{\ylm}{Y_{lm}}
\DeclareMathOperator{\lse}{LSE}

\begin{document}

\title{Accuracy of Charge Densities in Electronic Structure Calculations}

\author{Moritz Gubler} \email{moritz.gubler@unibas.ch}          \affiliation{\UniBasel}
\author{Moritz R. Schäfer} \affiliation{\rub} \affiliation{\rccs}
\author{Jörg Behler} \affiliation{\rub} \affiliation{\rccs}
\author{Stefan Goedecker} \affiliation{\UniBasel}

\begin{abstract}
Accurate charge densities are essential for reliable electronic structure calculations because they significantly impact predictions of various chemical properties and in particular, according to the Hellmann-Feynman theorem, atomic forces. This study examines the accuracy of charge densities obtained from different DFT exchange-correlation functionals in comparison with coupled cluster calculations with single and double excitations. We find that modern DFT functionals can provide highly accurate charge densities, particularly in case of meta-GGA and hybrid functionals. In connection with Gaussian basis sets, it is necessary to use the largest basis sets available  to obtain densities that are nearly basis set error free. These findings highlight the importance of selecting appropriate computational methods for generating high-precision charge densities, which are for instance needed to generate reference data for training modern machine learned potentials.

\end{abstract}

\maketitle

\section{Introduction}\label{sec:intro}
Obtaining accurate charge densities is essential in any contetxt where quantum mechanical calculations are required.
According to the Hohenberg-Kohn theorem~\cite{ksdft1}, the charge density contains the same complete information as the much more complicated many-body wave function. 
In particular, it states that the ground state electron density uniquely determines the ground state and all its properties.
The Hellmann-Feynman theorem~\cite{hellmann_feynman} connects the charge density to the forces acting on the nuclei. Since integrating the scalar product between the forces and the atomic displacements gives energy differences, the charge density is in this way directly related to the 
energy differences. In addition forces are essential quantities in many types of simulations such as geometry relaxations and molecular dynamics. 


Beyond these theoretical principles, charge densities are modeled for example in charge equilibration schemes used in polarizable force fields, where the dynamic redistribution of charges is the key to determining molecular interactions, polarization effects, and reactivity in complex systems.
 
In recent years, machine learning potentials have become an important addition to \abinit{} methods, reproducing electronic structure calculations with high accuracy at only a fraction of the computational cost~\cite{P4885,P5673,P5793,P6102,P6121,P6131,P6112,P5366}. 
Many recent machine learning potentials~\cite{4g,fast_4g,init_cent,cent,cent2,Zubatyuk_21,xie_20,jacobson_22,P6666,P6612,P5933} include effects like electrostatics and long range charge transfer, which necessitates the ability to accurately model charge densities using \abinit{} methods. Notably, some approaches~\cite{sadeghi_tb,ceriotti_1,ceriotti_2,ceriotti_3} focus on directly modeling the electron density itself, using it for constructing transferable and interpretable potentials.
These advancements highlight the critical role of accurate charge densities.
Due to the high computational costs it is generally not feasible to compute reference data such as atomic Hirshfeld charges~\cite{hirshfeld} using highly accurate methods like coupled cluster with single and double excitations (CCSD)~\cite{ccsd,ccsd_2}. Instead, more approximate methods like DFT~\cite{ksdft1,ksdft2} are typically used, trading some accuracy for increased computational efficiency.
However, the accuracy of charge densities is strongly influenced by the choice of the exchange correlation functional and even the most accurate DFT functionals suffer from an electron self-interaction error~\cite{delocalization_challenge,perdew_sic,perdew_spourious,quadratic_delocalization} that impacts the accuracy of the charge density. 
\citet{perdew_straying} analyzed electron densities for isolated atoms produced by 128 different DFT functionals and observed that while these densities became progressively closer to the exact ones until the early 2000s, this trend was reversed with the advent of unconstrained functionals that prioritized empirical fitting over physical rigor.
Additionally, several studies~\cite{wasilewski_ne,rodney_correlation,sala_diff,grabowski_vxc} have analyzed atomic charge densities to identify and understand the error sources in DFT, focusing on the impact of functionals and electron self interaction errors.
Controlling the self interaction error in approximate forms of the exchange correlation functional is a highly active field of research, with many approaches aiming to remove this source of error~\cite{tawada_lrc,laricchia_embedding,londsdale_sie,zheng_scaling}.
By evaluating semi-local density functionals on Hartree-Fock densities for transition states, \citet{perdew_barriers} demonstrated that this approach improves the accuracy barrier heights obtained with DFT calculations.
They attributed this improvement not to a reduction in density-driven errors, but to the introduction of a large density-driven error that cancels the functional-driven error which is consistent with our findings.

In this paper, we benchmark widely distributed Gaussian basis sets in combination with various DFT functionals and compare the accuracy of the obtained charge densities of molecules with CCSD calculations using the PySCF code~\cite{pyscf}. 
CCSD with pertubative triple excitations is not used here because the pertubative energy correction does not affect the charge density.
Specifically, we assess the performance of different basis sets, including polarization consistent~\cite{gauss_pc} and correlation consistent\cite{ccpv} sets, and analyze the resulting densities in terms of the numerical stability of Hirshfeld charge partitioning.
Such comparisons are essential for evaluating the reliability of computational approaches and guiding the selection of appropriate methodologies~\cite{cottenier_delta,elephant,compar_1,compar_2}.

By comparing the charge densities across various functionals and computational methods, we aim to identify the best strategies for obtaining high-precision and converged charge densities. Our findings have broad implications for computational chemistry, particularly in applications that require highly accurate modeling of electron density distributions such as modern machine learning potentials.

\section{Methods}\label{sec:methods}

\subsection{Kohn-Sham density functional theory}



Kohn-Sham DFT can be categorized into different approximations for the exchange correlation functional. 
The most common used approximations are:
\begin{itemize}
    \item Local density approximation (LDA) where $\varepsilon_{xc}(\rho)$ is a functional of the electron density at the same point.
    \item Generalized gradient approximation (GGA) where $\varepsilon_{xc}(\rho, \nabla \rho)$ is a functional of the electron density and its gradient.
    \item meta GGA (mGGA) where $\varepsilon_{xc}(\rho, \nabla \rho, \sum_i (\nabla \phi_i^*) (\nabla \phi_i))$ is a functional of the electron density, its gradient and the kinetic energy density.
    \item Hybrid functionals use a linear combination of an exchange correlation functional (either LDA, GGA or mGGA) 
    and the Hartree-Fock exchange energy.
    \item Range-separated hybrid functionals, in which the $1 / \norm{\vb r - \vb r'}$ kernel in the exchange energy is screened by a parameter $\mu$. The screening is applied through the expression $\erf \left( {\mu \norm{\vb r - \vb r'}} \right) / \norm{\vb r - \vb r'}$, where $\erf$ denotes the error function.
\end{itemize}

\subsection{Gaussian basis sets}
The spherical symmetry of the nuclear potential motivates the following ansatz for a basis function centered on atom $\alpha$:
$ \exp  (-a_k \norm{\vb r - \vb r_\alpha}^2) \cdot \ylm(\theta_\alpha, \varphi_\alpha)$ 
where $\norm{\vb r - \vb r_\alpha}$ is the distance to atom $\alpha$, $\theta_\alpha$, $\varphi_\alpha$ are the polar and azimuthal angle of $\vb r - \vb r_\alpha$ respectively, $\ylm(\theta_\alpha, \varphi_\alpha)$ is a spherical harmonic function and the parameter $a_k$ controls the with of the Gaussian basis function. Orbitals can then be expressed as linear combinations of these basis functions.

In Gaussian basis sets, the coefficients $a_k$ and the number of spherical harmonic functions $\ylm$ are chosen for each element by fitting these coefficients to reference values that are obtained analytically or with a more accurate basis set.

The biggest advantage of Gaussian orbitals is that overlap, kinetic, nuclear repulsion and Coulomb integrals 
can be solved analytically.
Disadvantages of Gaussian basis sets are that a large number of Gaussian functions is required to accurately represent the nuclear cusps in wave functions and Gaussian basis sets are not orthonormal meaning that convergence of the basis set to the complete basis set limit can not necessarily be achieved by adding more basis functions.
Finally, the condition number of the overlap matrix increases with the addition of Gaussian basis functions, leading to numerical problems that often limit the number of basis functions that can be used in a calculation.

\begin{figure}
    \centering
    \includegraphics{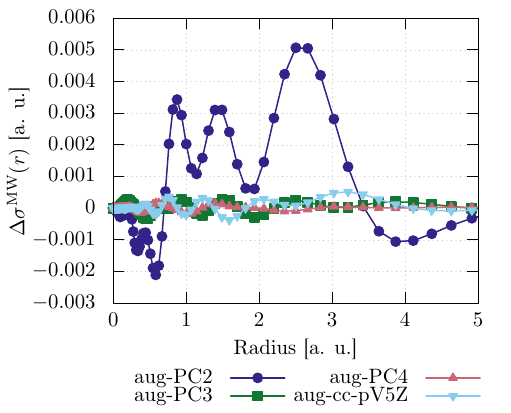}
    \caption{Basis set error $\Delta \sigma^{\mathrm{MW}}$ of DFT charge densities using the PBE functional for various DFT optimized, polarization consistent (PC) basis sets for a neon atom. A multi wavelet (MW) charge density is used as the reference value.}
    \label{fig:pc_conv}
\end{figure}

\subsection{Multi resolution wavelets}

Multi wavelets~\cite{Alpert.1993,Alpert.2002} provide a robust framework for representing quantum mechanical orbitals by utilizing adaptive grids and rigorous error control.
They form a systematic basis, which means that by adding more and more basis functions an arbitrarily high accuracy can be  achieved without running into numerical instabilities.
Unlike scalar wavelets, which rely on a single basis function per grid interval, multi wavelets use multiple basis functions inside each grid interval.

When a function is represented using multi wavelets, all quantities are mapped on an adaptive grid. Inside each grid interval multiple Legendre polynomials are used to represent the desired function. 
These Legendre polynomials are confined strictly within each grid element and are truncated at the edges resulting in discontinuities of the wave functions and other quantities at the borders of neighboring grid intervals. The fact that the truncation removes all overlap between multi wavelets across different grid intervals enables the efficient implementation of adaptive grids. The discontinuities between grid intervals do not impact the precision of integral calculations, which is why an integral based formulation has to be used when solving the Kohn–Sham equations in a multi-wavelet basis~\cite{mw_dft,mw_hf}.

The adaptivity of multi wavelets allows performing all-electron calculations where the resolution is increased for rapidly varying core orbitals, focusing computational resources where needed.
Rigorous error control within multi wavelet methods enables the user to define the desired precision before the calculation begins, ensuring that the basis set error remains smaller than the chosen precision.
For low precision wavelet calculations are generally slower than those using Gaussian orbitals, but when a large number of Gaussians is employed to obtain high precision, their efficiency becomes comparable. 
These features make multi wavelets particularly well suited for reference calculations in quantum mechanical applications, where both accuracy and efficiency are essential.

The multi wavelet approach can be coupled by a separable form to represent Greens functions of the Poisson and Helmholtz equation~\cite{beylkin_operator} which enables efficient Hartree-Fock~\cite{mw_hf} and DFT~\cite{mw_dft} calculations.
These features are implemented in the software package MRChem~\cite{mrchem,gubler_noise} and this code will be used for all wavelet based calculations in this paper.

\subsection{Accuracy measures of charge densities}\label{sec:compar_rho}

The charge densities for a variety of different exchange correlation functionals were compared with accurate coupled cluster calculations with single and double excitations (CCSD). PySCF~\cite{pyscf} was used for the CCSD calculations and the DFT calculation employing Gaussian basis sets.
The charge densities of twelve different DFT functionals and Hartree-Fock are benchmarked against CCSD charge densities for several molecules that are small enough to allow a CCSD calculation with an aug-cc-pV5Z basis set. 
In \cref{sec:bse_rho}, the precision of different Gaussian basis sets is compared and the choice of the aug-cc-V5Z basis set is justified.
The molecules are:
\begin{table}[H]
\centering
\begin{tabular}{llllll}
\ce{AlH3} & \ce{Ar} & \ce{BeH2} & \ce{BH3} & \ce{CH4} & \ce{F2} \\
\ce{H2} & \ce{H2O} & \ce{H2S} & \ce{HCl} & \ce{He} & \ce{HF} \\
\ce{LiH} & \ce{Li2} & \ce{MgH2} & \ce{N2} & \ce{NaH} & \ce{Ne} \\
\ce{NH3} & \ce{PH3} & \ce{SiH4} & \ce{CH2O} & \ce{CO2} & \ce{CO} \\
\ce{H2O2} & \ce{HCN} & & & & \\ 
\end{tabular}
\end{table}

For these molecules, six different measures of error with respect to the reference CCSD charge densities are calculated:
\begin{enumerate}
    \item Average difference of the Hirshfeld charges
    \item Average absolute value of maximal point wise charge difference.
    \item The average integral over $(\rho - \rho_{\text{CCSD}})^2$
    \item Average Coulomb energy of $\rho - \rho_{\text{CCSD}}$. This attenuates the short wavelength component of $(\rho - \rho_{\text{CCSD}})^2$.
    \item Average component-wise difference in dipole moments
    \item Average component-wise difference in quadrupole moments
\end{enumerate}
The integrals required to calculate the norms were evaluated numerically with the integration grids of PySCF using the accurate grid level. The spherically symmetric reference charge densities required for the calculation of the Hirshfeld charges were computed using Abinit~\cite{abinit}.

Measures 2–4 represent norms of the charge density error, where a value of 0 indicates a perfect match and higher values indicate larger errors. Measures 1, 5, and 6 are physically motivated and are best interpreted by comparing the results across the different functionals.
To illustrate the significance of deviations in Hirshfeld charges, consider a hydrogen molecule with a bond length of 1.4 Bohr. A change in the Hirshfeld charge by $2.4 \cdot 10^{-3}$ would correspond to an electron-nucleus interaction energy shift equal to the chemical accuracy threshold of 1 kcal/mol.
As a result, even relatively small changes in Hirshfeld charges can significantly impact the total energy.

In \cref{fig:dat_compare}, these measures are shown for all functionals.
The functionals tested are:
\begin{enumerate}
    \item[] LDA functionals
    \vspace{1ex}
    \hrule
    \item LDA with Slater exchange~\cite{slater_x} and VWN correlation~\cite{vwn_c}
    \item[] GGA functionals
    \vspace{1ex}
    \hrule
    \item PBE~\cite{pbe}
    \item r-PBE~\cite{rpbe}
    \item Becke exchange~\cite{becke_exchange} and LYP correlation~\cite{lyp_correlation} (BLYP)
    \item[] Hybrid GGA functionals
    \vspace{1ex}
    \hrule
    \item B3LYP~\cite{b3lyp_1,becke_exchange,b3lyp_2,b3lyp_3}
    \item PBE0~\cite{pbe0_1,pbe0_2}
    \item[] mGGA functionals
    \vspace{1ex}
    \hrule
    \item SCAN~\cite{scan}
    \item r-SCAN~\cite{rscan}
    \item r2-SCAN~\cite{r2scan}
    \item[] Hybrid, mGGA functionals
    \vspace{1ex}
    \hrule
    \item SCAN0~\cite{scan0}
    \item[] Range separated hybrid functionals
    \vspace{1ex}
    \hrule
    \item CAM-B3LYP~\cite{cam-b3lyp}
    \item WB97X-V~\cite{wb97,wb97x-v}
\end{enumerate}

\begin{figure}
    \centering
    \includegraphics{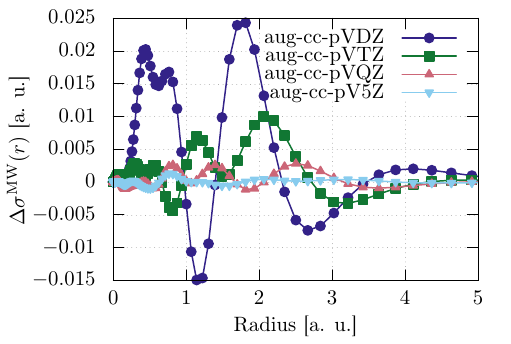}
    \caption{Basis set error $\Delta \sigma$ of DFT charge densities using the PBE functional for correlation consistent basis sets for a neon atom. A multi wavelet charge density is used as the reference value.}
    \label{fig:conv_ccpv}
\end{figure}

\begin{figure}
    \centering
    \includegraphics{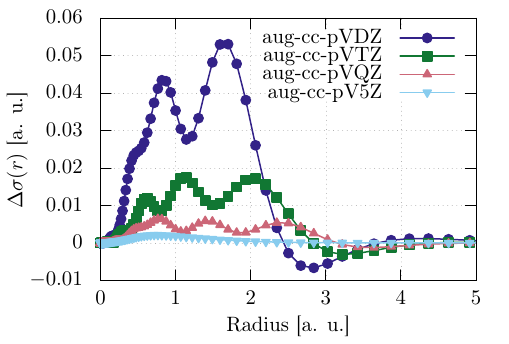}
    \caption{Basis set error $\Delta \sigma$ of CCSD charge densities with respect to the basis set. A calculation with a aug-cc-pwPV5Z was used as the reference charge density.}
    \label{fig:conv_ccsd}
\end{figure}

\begin{figure}
    \centering
    \includegraphics{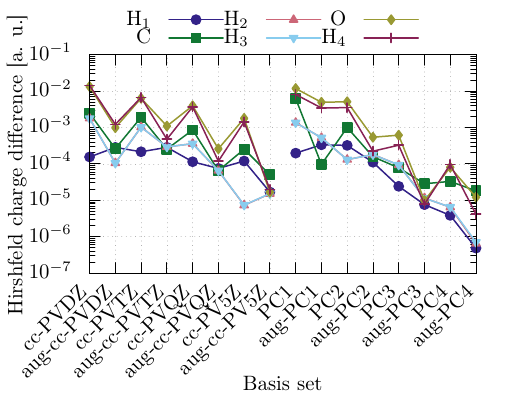}
    \caption{Basis set errors of Hirshfeld charges for different Gaussian basis sets for the atoms of a methanol molecule. Reference values were obtained using a multi wavelet calculation using the MRChem code. The same spherically symmetric reference charge densities were used to compute the Hirshfeld charges in all the different codes.}
    \label{fig:hf_methanol_error}
\end{figure}

\subsection{Hirshfeld charges}

Assigning partial charges to atoms in molecules is an important task in computational chemistry, as these charges serve several purposes: they can be used as parameters in force fields and machine learned potentials, or to qualitatively analyze the electronic structure of the system under study.
In this work, Hirshfeld charges are used as a measure for the quality of a functional's charge density by calculating the deviation to  Hirshfeld charges obtained using CCSD reference calculations.
Hirshfeld charges are one of many methods~\cite{mulliken,bader,ddec6,esp,resp,lowdin,natural_charges} to partition the electron density into atomic partial charges.
The Hirshfeld charge~\cite{hirshfeld} $q_i$ of atom $i$ is defined as
\begin{equation} \label{eq:hirshfeld}
    q_i = Z_i +  \int \partition \rho(\vb r) d\vb r
\end{equation}
where $\partition = \frac{\rho_i(\vb r)}{\sum_j \rho_j(\vb r)}$ is the partitioning function of atom $i$ and $Z_i$ the atomic number of atom $i$. The atomic reference charge densities $\rho_i(\vb r)$ are spherically symmetric around atom $i$ and are obtained and tabulated from free-atom calculations for each element in the system. Hirshfeld charges have the property that if $\rho(\vb r)$ is a superposition of the atomic reference charge densities then $q_i = \int \rho_i(\vb r) d\vb r = 0$. Because of this property, Hirshfeld charges can be seen as a measure of the change of the charge density close to atom $i$ compared to the reference charge density $\rho_i$.


In rare situations, the partitioning function $\partition$ may be numerically unstable since for points far away from all nuclei, both the numerator and the denominator are vanishing due to the exponential asymptotic decay of the charge density. However, in order to obtain accurate Hirshfeld charges, the partitioning function times the charge density must be integrated over the entire space. Numerical integration of \cref{eq:hirshfeld} mandates the evaluation of the partitioning function in these critical regions.

Ratios of two very small  quantities can be evaluated numerically in a numerically stable way with 
the log-sum-exp-trick. The log-sum-exp (LSE) operation is defined as $\lse(x_1, \dots, x_n) = \ln \left( \sum_{i=1}^n \exp(x_i) \right)$. By shifting all components with $x^* = \max \{x_1, \dots, x_n\} $, the LSE can be evaluated accurately even if the range of $x_i$ is large: $\lse(x_1, \dots, x_n) = x^* + \ln \left( \sum_{i=1}^n \exp(x_i - x^*) \right)$.
Let $\tilde \rho_i (\vb r) = \ln (\rho_i(\vb r))$. Using the numerically stable LSE function, the partitioning function can be rewritten as

\begin{equation} \label{eq:stable_partitioning}
    \partition = \exp \left( \tilde \rho_i - \lse(\tilde \rho_1(\vb r), \dots, \tilde \rho_n(\vb r)) \right).
\end{equation}

The expression in \cref{eq:stable_partitioning} is numerically stable for all values of $\rho_i(\vb r)$. We implemented this in PySCF~\cite{pyscf} and the multi wavelet based software MRChem~\cite{mrchem,gubler_noise}.

\section{Results}
\subsection{Basis set errors of charge densities}\label{sec:bse_rho}
Before comparing charge densities obtained with different exchange correlation functionals, it is essential to assess the density error resulting from the Gaussian basis functions.
To do this, the basis set errors of DFT using the PBE functional~\cite{pbe} are compared against basis-set-error-free charge densities obtained from a MRChem~\cite{mrchem} multi wavelet calculation.

A single neon atom was chosen as a test system since its charge density is spherically symmetric which simplifies visualization and comparison of charge densities. In this context, a useful quantity is 
\begin{equation}
    \sigma(r) = \int_{r = 0}^r \int_{\phi = 0}^{2 \pi} \int_{\theta = 0}^{\pi} \rho(r', \theta, \phi) r'^2 \sin \theta dr' d\phi d\theta,
\end{equation}
which is a one dimensional function $\sigma(r)$ that contains the charge inside a sphere of radius $r$.
When spherically symmetric charge densities are considered, no information about the three-dimensional charge density is lost. 
When comparing the charge density obtained with a DFT functional and a CCSD charge density, plotting the difference in $\sigma$ offers insights into the spatial error distribution of the charge density. The difference in $\sigma$ between two methods has units of charge and is the difference of charge contained in a sphere with radius $r$. It is called $\Delta \sigma^{\text{CC}}$ when a DFT charge density is compared to CCSD results and $\Delta \sigma^{\text{MW}}$ when a DFT charge density that was obtained using a Gaussian basis set is compared with a multi wavelet calculation in this paper.

First, the error of the charge density for polarization consistent Gaussian basis sets~\cite{gauss_pc} that are optimized for DFT calculations is analyzed.
In \cref{fig:pc_conv}, the difference of $\sigma$ between a basis-set-error-free multi wavelet charge density and a charge density obtained with polarization consistent basis sets~\cite{gauss_pc} is shown. The charge density was calculated with the PBE functional for a single neon atom. The ``aug'' prefix indicates the addition of diffuse functions in the Gaussian basis set.
The aug-PC3 and aug-PC4 basis sets agree well with the multi wavelet reference calculation. For comparison, a charge density that was obtained using a correlation consistent basis set (aug-cc-pV5Z) is added in \cref{fig:pc_conv} which gives the same level of accuracy in the charge density. The aug-PC3 contains fewer basis functions than the aug-cc-pV5z basis set, the aug-PC4 more.
In \cref{fig:conv_ccpv} the basis set error of charge densities obtained with a correlation consistent basis~\cite{ccpv} sets is shown.
Also, the aug-cc-pV5Z basis set matches the multi wavelet result almost perfectly. It is therefore recommended using this basis set whenever highly accurate charge densities are necessary with wave function based methods.
The aug-PC3 and aug-PC4 offer the same level with a similar number of basis functions, the aug-PC3 is even a bit smaller than the aug-cc-pV5Z and is therefore an interesting alternative to the correlation consistent basis set when DFT is used.
In \cref{fig:conv_ccsd} it is shown that the aug-cc-pV5Z basis set also produces highly accurate charge densities for coupled cluster calculations with single and double excitations.

To investigate the basis set error of charge densities for a system without spherical symmetry, Hirshfeld charges are analyzed in \cref{fig:hf_methanol_error} for a methanol molecule. The reference charge density was obtained using a multi wavelet DFT calculation with the PBE functional. As a measure of error, in this case the absolute value of the difference in Hirshfeld charges is used. Here, the addition of diffuse Gaussian basis functions greatly improves the accuracy and once again, the aug-cc-pV5Z basis is able to produce highly accurate charge densities.
The aug-PC3 and aug-PC4 basis sets also produce highly converged Hirshfeld charges, the aug-PC4 appears to be the most accurate basis set in this comparison. However, the calculation using the aug-PC4 basis set was difficult to converge since the condition number of the overlap matrix from the basis functions was larger than $10^8$.
The aug-PC3, aug-PC4 and the aug-cc-pV5Z bases offer sufficient precision to accurately compare charge densities obtained with different functionals.

\begin{figure*}
    \centering
    \includegraphics{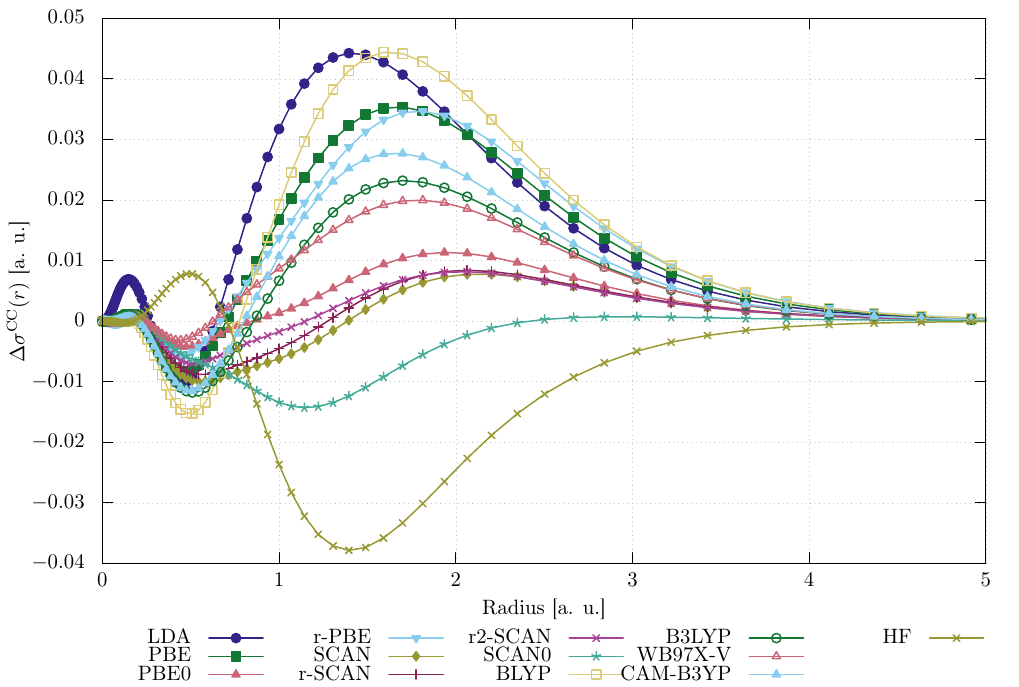}
    \caption{The difference in $\sigma(r)$ for CCSD charge densities and various DFT functionals of a neon atom. Negative values mean that the charge density of a functional contains more electrons than the CCSD reference calculation for a given radius.}
    \label{fig:atomic_methods}
\end{figure*}

Given that DFT results are being compared to CCSD calculations, it is more consistent to use the correlation-consistent aug-cc-pV5Z basis set, which is specifically designed for post-Hartree-Fock methods.

In Fig. 4 of our recent work on accurate force calculations using multi-wavelets~\cite{gubler_noise}, we analyzed the energy and force convergence for the same methanol molecule used in this study to compute Hirshfeld charges. That analysis showed that achieving chemical accuracy requires at least a quadruple-zeta basis set. To ensure an additional margin of accuracy, we opted for a quintuple-zeta basis set.
Moreover, the earlier study demonstrated that augmented basis functions significantly improve the accuracy of forces. A trend similarly observed for the Hirshfeld charges in this work. 
Therefore, in the remainder of this paper, the aug-cc-pV5Z basis set will be used for comparing various exchange-correlation functionals.

\begin{figure*}
\centering  
    \subfloat{
        \includegraphics{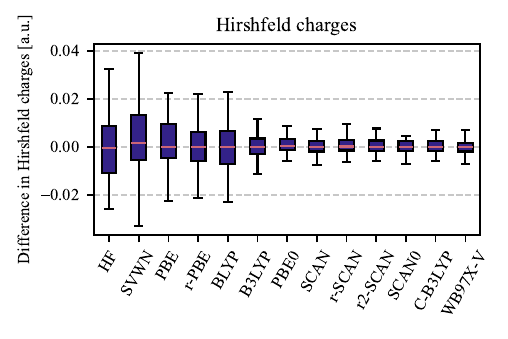}
    }
    \subfloat{
        \includegraphics{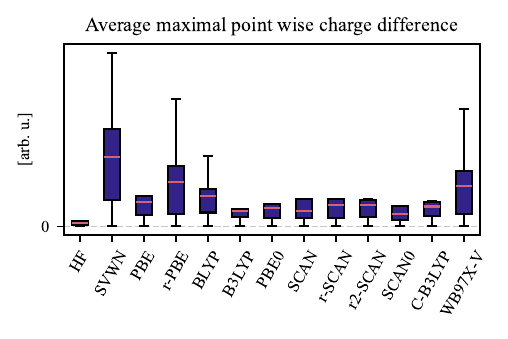}
    }
    
    \subfloat{
        \includegraphics{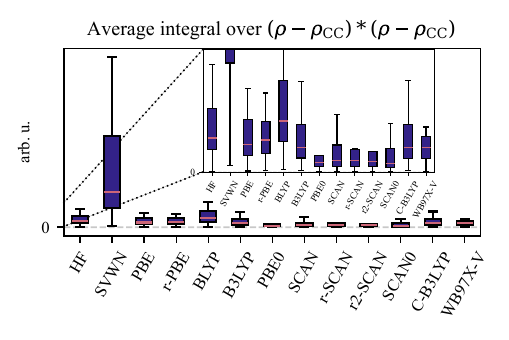}
    }
    \subfloat{
        \includegraphics{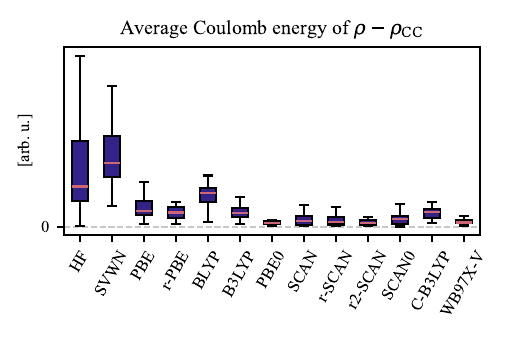}
    }

    \subfloat{
        \includegraphics{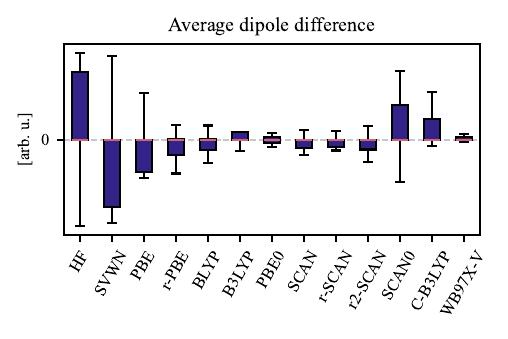}
    }
    \subfloat{
        \includegraphics{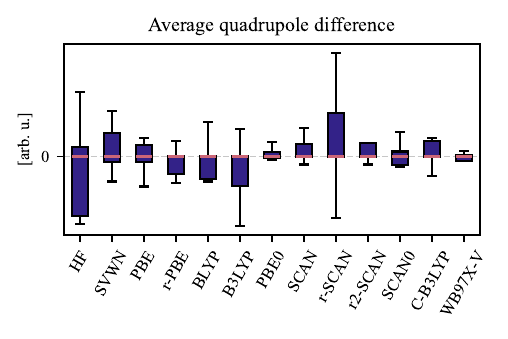}
    }
    
    \caption{
    Average errors of all measures for the quality of charge densities for a series of benchmark molecules. The calculations were done with the Gaussian aug-cc-pV5Z basis set and the PySCF code. CCSD charge densities were used as reference charge densities to compute the error of a given method. The CAM-BLYP functional is abbreviated with C-B3LYP.
    The box ranges from the first to the third quartile of the data, with the median indicated by a pink line. The whiskers extend from the edges of the box to the furthest data points within 1.5 times the interquartile range.
    }
    \label{fig:dat_compare}
\end{figure*}

\subsection{Accuracy of different exchange correlation functionals}

We now analyze the accuracy of the different exchange correlation functionals for the investigated set of small molecules.
Our previously defined measures of accuracy of the charge densities can be divided in two categories: point-wise measures and integral-based measures. The only point-wise measure used is the average maximal difference in charge density. The maximal difference in charge density is always located at the nuclear cusps. In this point-wise comparison, the Hartree-Fock charge density most accurately approximates the CCSD charge density. The LDA charge density is worst regarding the point-wise measure and there is no clear trend in the remaining DFT functionals.

However, the situation changes when considering the integral-based measures. In this case, HF and LDA yield poor charge densities, while all other functionals show significantly higher accuracy.
The PBE, rPBE, and the BLYP functional are able to reproduce CCSD charge densities accurately in most tests, but they struggle to accurately predict good dipole and quadrupole moments. The functionals that consistently produce good charge densities all either hybrid or meta-GGA. There is no clear benefit in the usage of range-separated hybrid functionals compared to modern state-of-the-art meta-GGA functionals such as r2-SCAN. The accuracy in the charge densities is similar in meta-GGA and hybrid functionals, making meta-GGA functionals the optimal choice that balance accuracy in charge densities and computational cost.
Moreover, exact exchange is not needed to compute accurate Hirshfeld charges. The SCAN, r-SCAN and r2-SCAN functionals produce highly accurate Hirshfeld charges with an average relative error of around 4 percent making them ideal methods to compute reference data for machine learning potentials. B3LYP, PBE0, SCAN0 and the range-separated hybrid functionals CAM-B3LYP and WB97X-V are similarly accurate. Since they all contain exact exchange they are computationally significantly more expensive.
It is noteworthy that all empirically designed functionals exhibit a significantly larger spread in at least one error measure compared to the other methods.
In contrast, the strongly constrained SCAN and r2-SCAN functionals, which satisfy all known meta-GGA functional constraints, consistently show less variation.
The r-SCAN functional exhibits a significantly larger spread in the average quadrupole difference compared to the other functionals, possibly due to its violation of some known meta-GGA constraints~\cite{r2scan}.

Very similar trends are found when analyzing a single atom. In \cref{fig:atomic_methods}, the difference in $\sigma(r)$ is shown for the tested functionals. It is found that r2-SCAN, SCAN and SCAN0 have the smallest error closely followed by PBE0, B3LYP and the range-separated hybrid functional WB97X-V.

\section{Conclusion}

A thorough investigation of the charge density error of Gaussian basis sets revealed that a very large basis set is required to accurately model the charge density. Polarization consistent basis sets that have been originally optimized for DFT calculations contain lower angular momenta in the Gaussian basis sets compared to the correlation consistent basis sets optimized for post-Hartree-Fock calculations.
The aug-cc-pV5Z, aug-PC3 and aug-PC4 basis sets are able to obtain accurate DFT charge densities making them the optimal choice for situations where highly accurate charge densities are needed.
When molecular systems are considered, adding diffuse basis functions greatly improves the accuracy of the charge density. The aug-cc-pV5Z basis set appears to be accurate enough to obtain good charge densities. Our results are also of interest in the context of constructing machine learned potentials whose atomic charges are fitted to density functional results~\cite{4g}. 
Since the atomic Hirshfeld charges 
obtained by different exchange correlation functionals 
differ by a few percent, there is no point trying to fit them 
in machine learning schemes to much higher precision.

An analysis of various metrics assessing the charge density quality showed that modern DFT functionals outperform Hartree-Fock for simple, mostly sp-bonded molecules at their equilibrium geometries.
For systems containing stretched bonds, or orbitals with higher angular momenta such as d and f orbitals, further benchmarking will be required. Conclusions from this analysis are therefore valid for similar systems.
The only functional that performs worse than HF is LDA. In contrast, all GGA, meta-GGA, and hybrid functionals evaluated in this study yield more accurate charge densities than HF. This trend suggests that the modern variants of meta-GGA and hybrid functionals are continuously enhancing the accuracy of charge densities in electronic structure calculations.
The sole measure where HF excels is the maximal point-wise error in the charge density, which consistently occurs at the nuclear positions.

Notably, the functionals that adhere to theoretical constraints, such as SCAN and r2-SCAN, produced the most consistent and accurate results across all error measures.
In contrast, empirically designed functionals exhibited a significantly larger spread in at least one error measure, emphasizing the importance of adhering to physically rigorous constraints for achieving consistently accurate charge densities.


\section{Code Availability} \label{sec:code}
A modified copy of MRCHEM that allows the calculation of Hirshfeld charges can be found in the GitHub repository: \url{https://github.com/moritzgubler/mrchem/tree/property/hirshfeld}. A python library was developed to analyze and compare DFT and CCSD charge densities using the PySCF~\cite{pyscf} framework. This library can be found here: \url{https://github.com/moritzgubler/charge-partitioning}

\begin{acknowledgements}
 Financial support was obtained from the Swiss National Science Foundation (project 200021 191994) and the Deutsche Forschungsgemeinschaft (DFG, German Research Foundation, project 495842446) in the framework of the DFG priority program SPP 2363. The calculations were performed on the computational resources of the Swiss National Supercomputer (CSCS) under project s1167 and on the Scicore (\url{http://scicore.unibas.ch/}) computing center of the University of Basel.
\end{acknowledgements}

\bibliographystyle{apsrev4-2}
\bibliography{main}

\end{document}